\documentstyle[12pt,epsf]{article}
\def\fun#1#2{\lower3.6pt\vbox{\baselineskip0pt\lineskip.9pt
\ialign{$\mathsurround=0pt#1\hfil##\hfil$\crcr#2\crcr\sim\crcr}}}

\newcommand{\be}{\begin{eqnarray}}
\newcommand{\ee}{\end{eqnarray}}
\newcommand{\bd}{\begin{displaymath}}
\newcommand{\ed}{\end{displaymath}}
\newcommand{\ba}{\begin{array}}
\newcommand{\ea}{\end{array}}
\newcommand{\bt}{\begin{tabular}}
\newcommand{\et}{\end{tabular}}

\begin{document}

\begin{flushright}
INLO-PUB-6/98\\
ITEP-TH-24/98
\end{flushright}

\vspace{.5cm}

\begin{center}
{\large
Isolated vacua in supersymmetric Yang--Mills theories}

\vspace{.5cm}

 {\bf A. Keurentjes} \\
{\it Instituut-Lorentz, University of Leiden, PO Box 9506, 2300 RA Leiden, 
The Netherlands}\\ {\bf A. Rosly} and  {\bf A.V. Smilga}\\
{\it ITEP, B. Cheremushkinskaya 25, Moscow 117218, Russia} \\

\end{center}

\vspace{.5cm}

\begin{abstract}
An explicit proof of the existence of  nontrivial  vacua in the pure 
supersymmetric Yang--Mills theories with higher orthogonal $SO(N), N \geq 7$
 or the $G_2$ gauge group defined on a 3--torus with periodic boundary 
conditions is given. Extra vacuum states are separated by an energy barrier 
from the perturbative vacuum $A_i = 0$ and its gauge copies.
\end{abstract}

It was shown by Witten long time ago \cite{IWit} that, in a pure $N=1$ 
supersymmetric gauge theory with any simple gauge group, the supersymmetry 
is not broken spontaneously. Placing the theory in a finite spatial box, 
the number of supersymmetric vacuum states [the Witten index ${\rm Tr} (-1)
^F$ ] was calculated to be  ${\rm Tr} (-1)^F = r + 1$ where $r$ is the rank 
of the gauge group. This results conforms with other estimates for 
 ${\rm Tr} (-1)^F$ for unitary and symplectic groups. It {\it disagrees}, 
however, with the general result 
\footnote{which follows e.g. from the counting of gluino zero modes on the 
instanton background \cite{IWit,IDyn}}
  \be
 \label{IW}
 {\rm Tr} (-1)^F \ =\ T(G)
 \ee
($T(G)$ is the Dynkin index of the adjoint representation) for higher 
orthogonal and exceptional groups. For $SO(N \geq 7)$, $T(G) = N-2  > r+1$.
Also for exceptional groups $G_2, F_4, E_{6,7,8}$, the index (\ref{IW}) is 
larger  than Witten's original estimate.

This paradox persisting for more than 15 years has been recently resolved by 
Witten himself \cite{Witnew}. He has found a flaw in his original arguments 
and 
shown that, for $SO(N \geq 7)$, vacuum moduli space is richer than it was 
thought before so that the {\it total} number  of quantum vacua is $N-2$ in 
accordance with the result (\ref{IW}). This note presents basically a 
comment to Witten's recent paper. Its {\it raison d'\^{e}tre} is to derive 
this result in an explicit way and in a form understandable to 
pedestrians (the paper \cite{Witnew} is full of special mathematical 
terminology and concepts which makes it difficult to understand for a person 
not familiar with this language). We also extend the analysis to the $G_2$ 
gauge group.

Let us first recall briefly Witten's original reasoning. 
 \begin{itemize}
\item Put our theory on the spatial 3D torus and impose {\it periodic} 
boundary conditions on the gauge fields \footnote{For unitary groups, 
one can perform the counting also with 't Hooft twisted boundary conditions, 
but for the orthogonal and exceptional groups where the mismatch in the 
Witten index calculations was observed this method does not work.}. Choose 
the gauge $A_0^a = 0$. A classical vacuum is defined as a gauge field 
configuration $A_i^a(x,y,z)$ with zero field strength (a {\it flat 
connection} in mathematical language).
 \item For any flat periodic connection, we can pick out a particular point 
in our torus $(0,0,0) \equiv (L,0,0) \equiv \ldots$ and define a set of 
holonomies (Wilson loops along nontrivial cycles of the torus)
 \be
 \label{hol}
\Omega_1 \ =\ P \exp \left\{ i \int_0^L A_1(x,0,0)  dx \right\}
\nonumber \\ 
\Omega_2 \ =\ P \exp \left\{ i \int_0^L A_2(0,y,0)  dy \right\}
\\ 
\Omega_3 \ =\ P \exp \left\{ i \int_0^L A_3(0,0,z)  dz \right\}
\nonumber 
  \ee
( $A_i = A_i^a T^a$ where $T^a$ are the group generators in a given 
representation).
${\rm Tr} \{\Omega_i\}$ are invariant under periodic gauge transformations. 
 \item A necessary condition for the connection to be flat is that all the 
holonomies (\ref{hol}) commute $[\Omega_i, \Omega_j] = 0$. We will see 
shortly that, for a simple connected group with $\pi_1(G) = 0$, it is also
a sufficient condition for a flat periodic connection with given holonomies 
to exist.
  \item A sufficient condition for the group matrices to commute is that
their {\it logarithms} belong to a Cartan subalgebra of the corresponding 
Lie algebra. For unitary and simplectic groups, this is also a necessary 
condition. In other words, any set of commuting group matrices $\Omega_i$ 
with $[\Omega_i, \Omega_j] = 0$ can be presented in the form
 \be
\label{Cart}
\Omega_i \ =\ \exp\{iC_i\} ,\ \ \ \ [C_i, C_j] = 0
 \ee
A flat connection with the holonomies $\Omega_i$ is then just $A_i = C_i/L$.
 Witten's original assumption which 
came out not to be true is that this is also the case for all other groups.
Assuming this, Witten constructed an effective Born--Oppenheimer 
hamiltonian for the slow variables $A_i^a$. It involves $3r$ bosonic degrees 
of freedom ($r$ is the rank of the group) and their fermionic counterparts. 
Imposing further the condition of the Weyl symmetry (a remnant of the 
original gauge symmetry) for the eigenstates of 
this hamiltonian, one finds $r+1$ supersymmetric quantum vacuum states.
 \end{itemize}

Before proceeding further, let us prove a simple

{\bf Theorem}: {\it For any set $\{\Omega_1, \Omega_2, \Omega_3\}, \ 
\Omega_i \in G$ where $G$ is a simple, connected, and simply connected 
group, $[\Omega_i, \Omega_j] = 0 \ \forall \, i,j$, a periodic flat 
connection exists such that $\Omega_i$ are the holonomies (\ref{hol}).}

\underline{Proof}: 
A flat connection (a pure gauge configuration) can be presented in the form
$A_i = -i \partial_i U U^{-1}$ with $U(x,y,z) \in G$. 
Let us seek for $U(x,y,z)$ satisfying the following boundary conditions
\be
 \label{bcU}
U(x+L,y,z) \ = U(x,y,z)  \Omega_1 \nonumber \\
U(x,y+L,z) \ = U(x,y,z)  \Omega_2 \\
U(x,y,z+L) \ = U(x,y,z)  \Omega_3 \nonumber
 \ee
with constant commuting $\Omega_i$  (commutativity of $\Omega_i$ is 
important for the matrix $U$ to be uniquely defined). Then $A_i(x,
y,z)$ is periodic. If choosing $U(0,0,0) = 1$, the matrices 
 $\Omega_i$ are the holonomies (\ref{hol}).
 We construct 
the matrix $ U(x,y,z)$ in several steps. 
 \begin{itemize}
\item
At the first step, we define
 \be
U(x,0,0) \ =\ \exp\left\{i\pi T_1 \frac xL \right\} \nonumber \\
U(0,y,0) \ =\ \exp\left\{i\pi T_2 \frac yL \right\} \\
U(0,0,z) \ =\ \exp\left\{i\pi T_3 \frac zL \right\} \nonumber 
  \ee
where $\Omega_i = \exp\{i\pi T_i\}$ (The choice of $T_i$ once $\Omega_i$ 
are given is not unique, but it is irrelevant. Take {\it some} set of the 
logarithms of holonomies $\Omega_i$). Having done this, we can extend the 
construction over all other edges of the 3-cube so that the boundary 
conditions (\ref{bcU}) are fulfilled. For example, we define
 $$
U(L,y,0) \ =\ \exp\left\{i\pi T_2 \frac yL \right\} \Omega_1,\ \ \ \ \
U(x,L,0) \ =\ \exp\left\{i\pi T_1 \frac xL \right\} \Omega_2 
 $$
etc.

\item With $U(x,y,z)$ defined on the edges of the cube in hand, we can 
continue $U$ also to the {\it faces} of the cube due to the fact that, 
according to our assumption, $\pi_1(G) = 0$ i.e. any loop in the group is
contractible. Let us do this first for 3 faces adjacent to the vertex (0,0,0)
.
\item With $U(x,y,0),\ U(x,0,z)$, and $U(0,y,z)$ in hand, we can find 
$U(x,y,z)$ on the other 3 faces of the cube:
$$U(x,y,L)  = U(x,y,0) \Omega_3,\ \ U(x,L,z)  = U(x,0,z) \Omega_2, \ \ 
U(L,y,z)  = U(0,y,z) \Omega_1 $$

\item With $U(x,y,z)$ defined on the surface of the cube, we can continue 
it into the interior using the fact that  $\pi_2(G) = 0$ for all simple Lie groups.
 \end{itemize}

By construction, $U(x,y,z)$ satisfies the boundary conditions (\ref{bcU})
and hence $A_i(x,y,z)$ is periodic.

 The skeleton 
construction just outlined is rather common in homotopy theory and can be 
found also in physical literature 
(see e.g.  \cite{Baal}). The proof works only for 
simply connected groups. If $\pi_1(G)  \neq 0$, it is generally not true, 
i.e. not for every set of commuting $\Omega_i$ a flat periodic connection 
with the holonomies $\Omega_i$ exists. The simplest counterexample is the 
set of three $SO(3)$ matrices 
 \be
\label{O3}
\Omega_1 = {\rm diag}(1,-1,-1);\ \Omega_2 = {\rm diag}(-1,1,-1);\ 
\Omega_3 = {\rm diag}(-1,-1,1)  
 \ee
Being diagonal, they obviously commute, but no corresponding periodic flat 
connection can be constructed. Indeed, suppose we have an $SO(3)$ connection 
$A_i^a(x,y,z)$ with the  holonomies  (\ref{O3}). The functions
$A_i^a(x,y,z)$ can be thought of also as a connection corresponding to
 the covering group $SU(2)$. 
 Nothing prevents us then from constructing the holonomies in the 
fundamental $SU(2)$ representation, which must be the liftings of the 
holonomies (\ref{O3}) to the group $SU(2)$. 
 One can readily derive 
 \be
\label{SU2}
\Omega_1^{SU(2)} = \pm i\sigma_1,\ \Omega_2^{SU(2)} = \pm i\sigma_2,\ 
\Omega_3^{SU(2)} = \pm i\sigma_3
 \ee
But the matrices (\ref{SU2}) do {\it not} commute anymore, the logarithms 
of the products \\ $\Omega_1^{SU(2)} \Omega_2^{SU(2)} [\Omega_1^{SU(2)}]^{-1} [\Omega_2^{SU(2)}]^{-1}$, etc. define  nonzero fluxes of the magnetic field along the 
corresponding directions on the 3-torus. Thus, the connection 
 $A_i^a(x,y,z)$ cannot be flat.

In reference \cite{Witnew}, Witten constructs a set of 3 commuting $SO(7)$ 
matrices such that  
\begin{description}
\item{\it i)} they cannot be presented in the form $\Omega_i = \exp\{iC_i\}$
 with commuting $C_i$; and \item{\it ii)} the 
corresponding holonomies in the covering group Spin(7) do commute as well.
\end{description}

As the group Spin(7) [in contrast with $SO(7)$] is simply connected, the 
theorem which we have just proven guarantees the existence of the 
corresponding nontrivial periodic flat connection. A particular convenient 
choice of 3 commuting $SO(7)$ matrices is
 \be
\label{O7}
\ba{ @{}r @{} r @{} r @{} r @{} r @{} r @{} r @{} r  @ {} l}
 \Omega_1 \ =\ {\rm diag} (& 1,& -1,& -1,& -1,& 1,& 1,& -1 &) \\
 \Omega_2 \ =\ {\rm diag} (& 1,& -1,& 1,& 1,& -1,& -1,& -1 &) \\
 \Omega_3 \ =\ {\rm diag} (&-1,& 1,& 1,& -1,& 1,& -1,& -1 & ) \ea
  \ee
This set consists of seven ``rows'' with $\pm 1$ as the elements such that 
every one of the seven possible combinations like $\left( 
\begin{array}{@{} c @{}} 1\\1\\-1 \end{array} \right)$ involving at least one 
minus 
appears just once. As far as $SO(7)$ is concerned, many other choices of 
the set $\{\Omega_i\}$ differing from (\ref{O7}) by permutations of the 
rows is possible. They all can be obtained from each other by global $SO(7)
$ rotations. We have chosen a particular order of the rows anticipating the 
further $G_2$ applications.

Each $\Omega_i$ can be represented as an exponential of an $SO(7)$ generator. 
This representation is far from being unique. For example,
 \be
\label{log}
\Omega_1 \ =\ \exp \{i\pi[T_{34} - T_{27}]\} = 
\ \exp \{i\pi[T_{23} - T_{47}]\}  \ = \ldots \nonumber \\
\Omega_2 \ =\ \exp \{i\pi[T_{56} + T_{27}]\} = 
\ \exp \{i\pi[T_{25} + T_{67}]\}  \ = \ldots \\
\Omega_3 \ =\ \exp \{i\pi[T_{16} - T_{47}]\} = 
\ \exp \{i\pi[T_{14} + T_{67}]\}  \ = \ldots \nonumber
  \ee
where $T_{ij}$ are the generators of the rotations in the $(ij)$ plane. The 
point is that one {\it can}not choose the ``logarithms'' of $\Omega_i$ so 
that all of them commute. Suppose one could, then write 
$\Omega_i = \exp \{i \pi S_i\}$. $[S_i, S_j] = 0 \ \forall \, i,j$ implies 
that $[S_i, \Omega_j] =0 \ \forall \, i,j$. But, as one can easily check, a 
matrix that commutes with all three $\Omega_j$ given has to be diagonal. The 
generators of $SO(7)$ however are antisymmetric, so no generator of $SO(7)$ 
commutes with all $\Omega_i$. This proves that the assumption 
$[S_i, S_j] = 0$ is wrong.

The new vacuum is isolated. This can be seen as follows:
Try to perturb the $\Omega_i$ 
\be \label{pert}
\Omega'_i = \Omega_i(1 + i\alpha_i^a T^a)
\ee
and require that
all $\Omega'_i$ still commute.
This implies the conditions
\bd
\alpha_i^a \Omega_i [T^a, \Omega_j] = \alpha_j^a \Omega_j [T^a, \Omega_i]  \ed
To solve these, we note that:
\begin{itemize}
\item With the given $\Omega^i$ and the standard basis  $T^a$ of the $so(7)$
Lie algebra, 
either $[\Omega^i, T^a] = 0$ or $\{ \Omega^i, T^a \} = 0$
\item $\alpha_i^a = 0$ if $[T^a, \Omega_i] = 0$ (since $[T^a, \Omega_j] \neq 0$ for some $i \neq j$)
\item  $\alpha_i^a = \alpha_j^a$ if both $[T^a, \Omega_i]\neq 0 $ and 
$[T^a, \Omega_j] \neq 0$ 
\end{itemize}
From these observations it follows that we can rewrite (\ref{pert}) as
\be  
\Omega'_i = \Omega_i + i\beta^a [\Omega_i, T^a]    
\ee
with $\beta^a$ independent of $i$. This is a global group rotation, and not a 
nontrivial deformation. The new vacuum does not admit deformations, and hence
 is isolated.

$SO(7)$ has a real 8--component spinor representation. Spinors are 
transformed under rotations. A set of all the corresponding $8 \times 8$ 
matrices is called the Spin(7) group. Two Spin(7) matrices differing by a 
sign correspond to one and the same $SO(7)$ matrix. Thereby a covering
${\rm Spin(7)} \ \stackrel{Z_2}{\to} \ SO(7)$ exists much analogous to the 
familiar covering   $ SU(2) \ \stackrel{Z_2}{\to} \ SO(3)$. 
Witten proves that the Spin(7) holonomies corresponding to the $SO(7)$ 
holonomies (\ref{O7}) commute using a powerful mathematical machinery 
involving notions like Stieffel--Whitney class etc. We will show this 
explicitly.

The generators of Spin(7) are 
\be
\label{TS}
T^S_{ij} \ =\ \frac 14 [\Gamma_i, \Gamma_j]
 \ee
where $\Gamma_i$ are 7--dimensional $\Gamma$--matrices satisfying the 
Clifford algebra $\Gamma_i \Gamma_j + \Gamma_j \Gamma_i = -2\delta_{ij}$.
One particular choice for the $\Gamma$--matrices is
  \be
  \Gamma_1 = i\sigma^2 \otimes \sigma^2 \otimes \sigma^2; \ \ 
  \Gamma_2 = -i \otimes \sigma^1 \otimes \sigma^2; \ \ 
  \Gamma_3 = -i \otimes \sigma^3 \otimes \sigma^2; \ \ 
  \Gamma_4 = i\sigma^1 \otimes \sigma^2 \otimes 1;  \nonumber \\
  \Gamma_5 = -i\sigma^3 \otimes \sigma^2 \otimes 1; \ \ 
  \Gamma_6 = -i\sigma^2 \otimes 1 \otimes \sigma^1; \ \ 
  \Gamma_7 = -i\sigma^2 \otimes 1 \otimes \sigma^3
  \ee
It is not difficult now to construct explicitly the generators of Spin(7) 
and exponentiate them as in Eq.(\ref{log}). It does not matter which 
particular representation for $\Omega_i$ is chosen, it affects only the 
overall sign. The set of the Spin(7) holonomies corresponding to the set
(\ref{O7}) of the $SO(7)$ holonomies is
  \be
 \label{Spin}
\Omega_1^{spin} = \pm \sigma^3 \otimes 1 \otimes \sigma^3;\ \ 
\Omega_2^{spin} = \pm \sigma^3 \otimes \sigma^3 \otimes 1;\ \ 
\Omega_3^{spin} = \pm \sigma^3 \otimes 1 \otimes 1
  \ee
It is easy to see that $[\Omega_i^{spin}, \Omega_j^{spin}] = 0$. As 
$\pi_1[\rm{Spin}(7)] = 0$, non--trivial periodic flat connections with the 
holonomies (\ref{Spin}, \ref{O7}) exist.

A similar statement for the $G_2$ gauge group can be obtained free of 
charge. The $G_2$ group can be defined as a subgroup of $SO(7)$ leaving 
invariant the combination $f_{ijk}Q^i P^j R^k$ where $Q^i, P^j, R^k$ are 3 
arbitrary 7--vectors and $f_{ijk}$ is a certain antisymmetric tensor. One 
particular convention for $f_{ijk}$ is
 \be
f_{165} = f_{341} = f_{523} = f_{271} = f_{673} = f_{475} = f_{246} =1
 \ee
and all other non-zero components are recovered using antisymmetry. It is easy 
to see now that the matrices (\ref{O7}) do belong to the $G_2$ subgroup of 
$SO(7)$. 
Another way to see the same is to define $G_2$ as a subgroup of Spin(7) 
leaving a particular spinor invariant. The matrices (\ref{Spin}) leave 
invariant the spinor $\eta \ = \left( \begin{array}{c} 1 \\ 0 \end{array} 
\right) \otimes  \left( \begin{array}{c} 1 \\ 0 \end{array} \right)
\otimes  \left( \begin{array}{c} 1 \\ 0 \end{array} \right) $ and hence belong 
to $G_2$ (and, incidentally, $f_{ijk} = \eta^T \Gamma_{[i} \Gamma_j 
\Gamma_{k]} \eta$).
As $G_2$ is simply connected,  $\pi_1(G_2) = 0$ \footnote{This can be seen 
from the fact that the unique compact simply-connected group with Lie-algebra 
$G_2$ has a trivial center (see e.g. \cite{pi1G2})} , 
one can apply our general theorem immediately 
and make sure thereby that a non-trivial periodic flat $G_2$ connection exists.
 This extra vacuum state together with $r_{G_2} + 1 = 3$ ``old'' states 
associated with the constant gauge potentials belonging to the Cartan 
subalgebra makes the total vacuum 
state counting in accordance with the result ${\rm Tr} (-1)^F 
= T(G_2) = 4$.

For $SO(7)$ and $G_2$, the new corner in the moduli space of classical 
vacua presents just a single point. The same is true for $SO(8)$: up to a 
global gauge transformation, any set of commuting $SO(8)$ matrices whose 
logarithms do not commute can be presented in the form $\Omega^{SO(8)}_i
= {\rm diag} (\Omega^{SO(7)}_i, 1)$ with $\Omega^{SO(7)}_i$ given by 
Eq.(\ref{O7}). Consider still higher orthogonal groups. Starting from $SO(9)$, 
an additional freedom appears associated with Cartan rotations in extra 
dimensions; any set $\Omega^{SO(N)}_i = {\rm diag} (\Omega^{SO(7)}_i, 
\omega^{SO(N-7)}_i )$ with logarithms of
$\omega^{SO(N-7)}_i$ belonging to the Cartan subalgebra of $SO(N-7)$ gives
rise to a nontrivial $SO(N)$ connection. The extra component of the moduli 
space is not an isolated point anymore, but presents a manifold. 
Its dimension is $3r_{SO(N-7)}$. There are $r_{SO(N-7)} + 1$ 
eigenstates of the corresponding Born--Oppenheimer hamiltonian. All together
 we have $(r_{SO(N)} + 1) + (r_{SO(N-7)} + 1) \ =\ N-2$ vacuum states 
\cite{Witnew} in accordance with the counting (\ref{IW}).

There is some subtlety for $SO(9)$, where the continuous unbroken symmetry 
group is $SO(2)$,  which is 
abelian.  The index for $SO(2)$-theory is 
${\rm Tr} (-1)^F = 0$, which seems to lead to a wrong answer for 
$SO(9)$-theory. This is resolved as follows. Apart from the continuous 
$SO(2)$, there are also some discrete symetries unbroken. An example of such
 a discrete symmetry is represented by the 
matrix ${\rm diag}(1,1,1,1,1,1,-1,1,-1)$. This matrix commutes with the 
holonomies ${\rm diag}(\Omega^{SO(7)}_i,1,1)$, and acts as ${\rm diag}(1,-1)$ 
in the unbroken $SO(2)$-subgroup. It is a gauge symmetry, so we have to 
demand 
invariance under this symmetry. In this way, the unbroken $SO(2)$ is enhanced 
to $O(2)$, and we need ${\rm Tr}(-1)^F$ for $O(2)$-theory, not $SO(2)$. To 
calculate the index we can simply repeat the analysis for $SO(2)$ from 
$\cite{IWit}$, with the requirement of invariance under the 
extra symmetry ${\rm diag}(1,-1)$. One finds that, of the four states 
mentioned in \cite{IWit}, the two states with one fermion are not invariant 
under the extra symmetry, while the two bosonic states 
(two fermions or none) are invariant. In this way we find ${\rm Tr}(-1)^F=2$ for $O(2)$-theory, in 
contrast to the zero result of $SO(2)$-theory. Hence for $SO(9)$ one finds 
$(r_{SO(9)} + 1) + 2 \ =\ 7$, the right number. 

In an analogous way one finds that for the higher orthogonal groups, the 
unbroken symmetry group is actually $O(N-7)$ (for the Spin groups it is 
Pin($N-7$), the double cover of $O(N-7)$). However, the extra symmetry does 
not affect the analysis in this case, and the previous results stay valid.
  
At first sight, starting from $N = 14$, a new corner in the moduli space 
associated with the matrices ${\rm diag} (\Omega^{SO(7)}_i, 
\Omega^{SO(7)}_i )$
might appear. This is not so, however. One can write explicitly
  \be
\label{log14}
\Omega_1^{SO(14)} \ =\ {\rm diag}(1,-1,-1,-1,1,1,-1; 1,-1,-1,-1,1,1,-1)\ =
\nonumber \\ \exp\{i\pi[T_{2,9} + T_{3,10} + T_{4,11} + T_{7,14}] \}
  \ee
and similarly for $\Omega_{2}, \Omega_3$. $\log \Omega_i^{SO(14)}$ defined 
according to the prescription (\ref{log14}) commute and belong to the 
Cartan subalgebra of $SO(14)$. So we obtain nothing new. Also for still higher
 groups nothing new happens. After possibly rotating away non-trivial 
$SO(14)$-blocks, as in the above, all the information about the holonomies 
will be contained in an $SO(p)$-
subgroup (with $p < 14$). Hence for $SO(N)$, there are never more than two 
components in the moduli space, no matter how large $N$ gets.

It is an interesting and still unresolved problem how to find a similar 
explicit construction revealing extra $T(G) - r_G - 1$ vacuum states for 
four other exceptional groups. The corresponding numbers are listed in the 
Table.

\vspace{.5cm}

\begin{tabular}{||l|c|c|c|c||} \hline
group G & $F_4$ & $E_6$ & $E_7$ & $E_8$ \\ \hline
r + 1    &  5 & 7 & 8 & 9 \\ \hline
 T(G) & 9 & 12 & 18 & 30 \\ \hline
mismatch & 4 & 5 & 10 & 21 \\ \hline
\end{tabular}

\vspace{.5cm}

\centerline{{\bf Table}: Vacuum counting for higher exceptional groups.}

\vspace{.3cm}

For $E_7$, $T(G) > 2(r+1)$ while, for $E_8$, $T(G) > 3(r+1)$. Assuming that 
each disconnected 
component might contribute not more than $r + 1$ states in the total 
counting, it suggests the presence of at least {\it three} disconnected 
components for $E_7$ and at least four components for $E_8$.

The presence of extra disconnected components in the vacuum moduli space 
implies the existence of classical solutions to the Yang--Mills equations of 
motion for the field living on $T^3 \otimes R_\tau$ which interpolate between 
trivial vacua with constant commuting $A_i $ at $\tau = -\infty$ and the 
non--trivial one at $\tau = \infty$. Indeed, one can start with the 
configuration
 \be
 \label{qinst}
 A_i^{\rm trial}(\tau, \vec{x}) \ =\ \frac {1 + \tanh (\mu \tau)}{2}
 A_i^{\rm isol.\ vac.} (\vec{x})
 \ee
 and then deform it with the boundary conditions at $\tau = \pm \infty$ 
 fixed so that the action be minimized. A
solution thus obtained should have a finite action $\sim 1/g^2$. 
Probably, it can be 
found only numerically in a way similar to how the toron--like Euclidean 
solutions for the $SU(2)$ gauge group with 't Hooft twisted boundary 
conditions were earlier found in Ref.\cite{Garcia}.

These new Euclidean solutions have nothing to do with conventional 
instantons: the latter interpolate between  trivial vacua and the vacua
with nonzero integer Chern--Simons number but with the same trivial 
holonomies. 
An interesting question is, what is the Chern--Simons number 
of our isolated vacuum. 

{\bf Acknowledgements}: We are indebted to A. Losev, A. Vainshtein, and P. 
van Baal for illuminating discussions. This work was supported in part by the 
RFBR--INTAS grants 
93-0166, 93-0283 and 94-2851, by the RFFI grants 96-02-18046 and 97-02-16131, 
by the grant 96-15-96455 for support of scientific schools, by the U.S. 
Civilian 
Research and Development Foundation under award \# RP2-132, and by the 
Schweizerischer National Fonds grant \# 7SUPJ048716.

\end{document}